\documentclass[12pt]{article}
\usepackage{graphicx}
\def\bra#1{\left\langle #1\right|}
\def\ket#1{\left| #1\right\rangle}
\newcommand{\bers}{\begin{eqnarray*}}
\newcommand{\eers}{\end{eqnarray*}}
\newcommand{\bt}{\begin{itemize}}
\newcommand{\et}{\end{itemize}}
\def\beq{\begin{equation}}
\def\eeq{\end{equation}}
\def\bea{\begin{eqnarray}}
\def\eea{\end{eqnarray}}
\def\nn{\nonumber}
\def\sss{\scriptscriptstyle}
\def\bd{B_d^0}
\def\bdbar{{\bar B}^0_d}
\def\bs{B_s^0}

\def\barp{{\raise.35ex\hbox
{${\sss (}$}}---{\raise.35ex\hbox{${\sss )}$}}}
\def\bdbarp{\hbox{$B_d$\kern-1.4em\raise1.4ex\hbox{\barp}}}
\def\bsbarp{\hbox{$B_s$\kern-1.4em\raise1.4ex\hbox{\barp}}}
\def\ks{K_{\sss S}}
\def\kbar{{\bar K}^0}
\def\kstarbar{{\bar K}^*}
\def\EK{E_{\sss K}}
\def\roughly#1{\mathrel{\raise.3ex\hbox
{$#1$\kern-.75em\lower1ex\hbox{$\sim$}}}}
\def\lsim{\roughly<}

\def\bra#1{\langle#1|}
\def\ket#1{|#1\rangle}
\def\Puc{{\cal P}_{uc}}
\def\Ptc{{\cal P}_{tc}}
\def\Ptwiduc{{\tilde{\cal P}}_{uc}}
\def\Ptwidtc{{\tilde{\cal P}}_{tc}}
\def\Ptwiducp{{\tilde{\cal P}}'_{uc}}
\def\Ptwidtcp{{\tilde{\cal P}}'_{tc}}
\def\ZI{Z_{\sss I}}
\def\ZR{Z_{\sss R}}
\def\fbdk{F_{B_d \to K}}
\def\fbsk{F_{B_s \to {\bar K}}}
\def\fdk{F_{{\bar D} \to K}}
\def\fdsk{F_{{\bar D}_s \to {\bar K}}}
\def\fbdkstar{F_{B_d \to K^*}}
\def\fbskstar{F_{B_s \to {\bar K}^*}}
\def\fdkstar{F_{{\bar D} \to K^*}}
\def\fdskstar{F_{{\bar D}_s \to {\bar K}^*}}

\def\jhep#1#2#3{{\it JHEP}\ {\bf #1}, #3 (#2)}

\def\npb#1#2#3{{\it Nucl.\ Phys.} {\bf B#1}, #3 (#2)}
\def\plb#1#2#3{{\it Phys.\ Lett.} {\bf #1B}, #3 (#2)}
\def\prd#1#2#3{{\it Phys.\ Rev.} {\bf D#1}, #3 (#2)}
\def\newprd#1#2#3{{\it Phys.\ Rev.} {\bf D#1}: #3 (#2)}

\def\prl#1#2#3{{\it Phys.\ Rev.\ Lett.} {\bf #1}, #3 (#2)}

\begin{document}

\begin{flushright}
UdeM-GPP-TH-01-87\\
\end{flushright}
\vskip0.5truecm

\begin{center}
  {\large \bf \centerline{$B \to K^{(*)} {\bar K}^{(*)}$ Decays: A New
      Method for Measuring the CP Phase $\alpha$}} \vspace*{1.0cm}
{\large Alakabha Datta\footnote{email: datta@lps.umontreal.ca} and
  David London\footnote{email: london@lps.umontreal.ca}} \vskip0.3cm
{\it Laboratoire Ren\'e J.-A. L\'evesque, Universit\'e de
  Montr\'eal,} \\
{\it C.P. 6128, succ.\ centre-ville, Montr\'eal, QC, Canada H3C 3J7} \\
\vskip0.3cm
\bigskip
(\today)
\vskip0.5cm
{\Large Abstract\\}
\vskip3truemm
\parbox[t]{\textwidth} {We present a new method for measuring the CP
phase $\alpha$. It requires the measurement of the pure penguin decays
$\bd(t) \to K^{(*)} {\bar K}^{(*)}$ and $\bs \to K^{(*)} {\bar
K}^{(*)}$. The method is quite clean: we estimate the theoretical
uncertainty to be at most 5\%. By applying the method to several
$K^{(*)} {\bar K}^{(*)}$ final states, $\alpha$ can be extracted with
a fourfold ambiguity. An additional assumption reduces this ambiguity
to twofold: $\{ \alpha, ~\alpha + \pi \}$. Since no $\pi^0$ detection
is needed, this method can be used at hadron colliders.}
\end{center}
\thispagestyle{empty}
\newpage
\setcounter{page}{1}
\textheight 23.0 true cm
\baselineskip=14pt

There is a great deal of excitement these days regarding CP violation
in the $B$ system. The latest measurements of the CP-violating phase
$\beta$ have now produced definitive evidence for CP violation outside
the kaon system \cite{betameas}:
\beq
\sin 2\beta = 0.79 \pm 0.12 ~.
\eeq
The ultimate goal of the study of CP-violating rate asymmetries in $B$
decays is to measure each of the interior angles of the unitarity
triangle \cite{CPreview}, $\alpha$, $\beta$ and $\gamma$. In this way
we will be able to test the standard model (SM) explanation of CP
violation. $B$-factories have obtained $\beta$ by measuring the CP
asymmetry in the ``gold-plated'' decay mode $\bd(t) \to J/\psi
\ks$. And many methods have been proposed for measuring, or putting
limits on, the CP phase $\gamma$ \cite{CPreview}.

On the other hand, to date there are only two clean techniques for the
extraction of $\alpha$, and each has its particular difficulties. In
the first method, one uses the CP asymmetry in $\bd(t)\to\pi^+\pi^-$
to obtain $\alpha$. However, in order to remove the penguin
``pollution,'' it is necessary to perform an isospin analysis of
$B\to\pi\pi$ decays \cite{isospin}, which includes the measurement of
$\bd\to\pi^0\pi^0$. Since the branching ratio for this decay is
expected to be quite small, it may be very difficult to obtain
$\alpha$ in this way. Second, one can use a Dalitz-plot analysis of
$\bd(t)\to\rho\pi\to\pi^+\pi^-\pi^0$ decays \cite{Dalitz}. The problem
here is that one must understand the continuum background to such
decays with considerable accuracy, as well as the correct description
of $\rho\to\pi\pi$ decays, and again these may be difficult. Note also
that both methods require the detection of $\pi^0$'s, which makes them
a challenge for hadron colliders.

In this paper, we present a new method for measuring $\alpha$ based on
the pure penguin decays $\bd(t) \to K^{(*)} {\bar K}^{(*)}$ and $\bs
\to K^{(*)} {\bar K}^{(*)}$, which are related by U-spin. By studying
these decays in the limit of heavy-quark symmetry, chiral symmetry,
and the large-energy limit of QCD for the final-state kaons, we argue
that the $SU(3)$-breaking effects are quite a bit smaller than what is
usually found. In particular, our best estimate of the theoretical
error in our method is at most 5\%, which makes the extraction of
$\alpha$ quite clean. Note that it is possible to make a variety of
independent experimental measurements which will test the claim of
small $SU(3)$ breaking.

Because the branching ratios for $\bd(t) \to K^{(*)} {\bar K}^{(*)}$
are rather small, and because $\bs$ decays are involved, this method
is probably most appropriate for hadron colliders, particularly since
no $\pi^0$ detection is needed. Still, it is not out of the question
that $e^+e^-$ $B$-factories might be able to use this technique. One
potential drawback of this method is the presence of multiple discrete
ambiguities. However, by combining information from several final
$K^{(*)} {\bar K}^{(*)}$ states, it is possible to reduce the
ambiguity in $\alpha$ to a fourfold one. And by imposing a further
(reasonable) theoretical condition, one can obtain only a twofold
ambiguity: $\{ \alpha, ~\alpha + \pi \}$.

Consider the pure $b\to d$ penguin decay $\bd \to K^0 \kbar$. At the
quark level, the decay takes the form ${\bar b} \to {\bar d} s {\bar
  s}$. The amplitude can be written
\bea
A(\bd\to K^0\kbar) & = & P_u V_{ub}^* V_{ud} + P_c V_{cb}^* V_{cd} +
P_t V_{tb}^* V_{td} \nn\\
& = & \Puc\ e^{i \gamma} e^{i \delta_{uc}} + \Ptc\ e^{-i \beta} e^{i
  \delta_{tc}} ~,
\label{BKKamp}
\eea
where $\Puc\equiv |(P_u - P_c) V_{ub}^* V_{ud}|$, $\Ptc\equiv |(P_t -
P_c) V_{tb}^* V_{td}|$, and we have explicitly written out the strong
phases $\delta_{uc}$ and $\delta_{tc}$, as well as the weak phases
$\beta$ and $\gamma$. (In passing from the first line to the second,
we have used the unitarity of the Cabibbo-Kobayashi-Maskawa (CKM)
matrix, $V_{ub}^* V_{ud} + V_{cb}^* V_{cd} + V_{tb}^* V_{td} = 0$, to
eliminate the $V_{cb}^* V_{cd}$ term.) The amplitude ${\bar A}$
describing the conjugate decay $\bdbar \to K^0 \kbar$ can be obtained
from the above by changing the signs of the weak phases.

By making time-dependent measurements of $\bd(t)\to K^0\kbar$, one can
obtain the three observables
\bea
X &\equiv & \frac{1}{2} \left( |A|^2 + |\bar{A}|^2 \right) =
\Puc^2 + \Ptc^2 - 2 \Puc \Ptc \cos\Delta \cos\alpha ~, \nn \\
Y &\equiv & \frac{1}{2} \left( |A|^2 - |\bar{A}|^2 \right) =
- 2 \Puc \Ptc \sin\Delta \sin\alpha ~, \\
\label{BKKobservables}
\ZI &\equiv & {\rm Im}\left( e^{-2i \beta} A^* {\bar A} \right)
= \Puc^2 \sin 2\alpha - 2 \Puc \Ptc \cos\Delta \sin\alpha ~, \nn
\eea
where ${\Delta}\equiv {\delta}_{uc} - {\delta}_{tc}$. It is useful to
define a fourth observable:
\bea
\ZR & \equiv & {\rm Re}\left( e^{-2i \beta} A^* \bar{A} \right) \\
\label{ZR}
& = & \Puc^2 \cos 2\alpha +\Ptc^2 - 2 \Puc \Ptc \cos\Delta \cos\alpha ~.
\nn
\eea
The quantity $\ZR$ is not independent of the other three observables:
\beq
\ZR^2 = X^2 - Y^2 - \ZI^2 ~.
\eeq
Thus, one can obtain $\ZR$ from measurements of $X$, $Y$ and $\ZI$, up
to a sign ambiguity. Note that the three independent observables
depend on four theoretical parameters ($\Puc$, $\Ptc$, $\Delta$,
$\alpha$), so that one cannot obtain CP phase information from these
measurements \cite{LSS}. However, one can partially solve the
equations to obtain
\beq
\Ptc^2 = {\ZR \cos 2\alpha + \ZI \sin 2\alpha - X \over \cos 2\alpha -
  1} ~.
\eeq

Now consider a second pure $b\to d$ penguin decay of the form $\bd \to
K^* \kstarbar$. Here $K^*$ represents any excited neutral kaon, such as
$K^*(892)$, $K_1(1270)$, etc. This second decay can be treated
completely analogously to the first one above, with unprimed
parameters and observables being replaced by primed ones. One can then
combine measurements of the two decays to obtain
\beq
{\Ptc^2 \over {\Ptc'}^2} = {\ZI \sin 2\alpha + \ZR \cos 2\alpha - X
  \over \ZI' \sin 2\alpha + \ZR' \cos 2\alpha - X' } ~.
\label{alphasolve}
\eeq

Now comes the main point. The ratio $\Ptc^2 / {\Ptc'}^2$ can be
obtained by measuring $\bs$ decays to the same final states $K^0\kbar$
and $K^* \kstarbar$. Consider first the decay $\bs\to K^0\kbar$. This
is described by a $b\to s$ penguin amplitude:
\bea
A(\bs\to K^0\kbar) & = & P_u^{(s)} V_{ub}^* V_{us} + P_c^{(s)}
V_{cb}^* V_{cs} + P_t^{(s)} V_{tb}^* V_{ts} \nn\\
& \simeq & (P_t^{(s)} - P_c^{(s)}) V_{tb}^* V_{ts} \equiv \Ptc^{(s)} ~.
\label{Bsdecay}
\eea
In writing the second line, we have again used the unitarity of the
CKM matrix to eliminate the $V_{cb}^* V_{cd}$ piece. Furthermore, the
$V_{ub}^* V_{us}$ piece is negligible: $|V_{ub}^* V_{us}| \ll
|V_{tb}^* V_{ts}|$. Thus, the measurement of the branching ratio for
$\bs\to K^0\kbar$ yields $|\Ptc^{(s)}|$. Similarly one can obtain
$|\Ptc^{'(s)}|$ from the branching ratio for $\bs \to K^* \kstarbar$.
However, to a very good approximation,
\beq
{{\Ptc^{(s)}}^2 \over {\Ptc^{'(s)}}^2} = {\Ptc^2 \over {\Ptc'}^2} ~.
\label{Pratios}
\eeq
(Note that the CKM matrix elements cancel in both ratios.) As we will
argue below, the theoretical error in making this approximation is at
most 5\%. The measurements of the branching ratios for $\bs\to
K^0\kbar$ and $\bs \to K^* \kstarbar$ will therefore allow one to
obtain $\Ptc^2 /{\Ptc'}^2$. Thus, by combining Eqs.~(\ref{alphasolve})
and (\ref{Pratios}), one can extract $\alpha$ quite cleanly (up to
discrete ambiguities, which will be discussed below).

A modification of this method can also be used when the final state is
not self-conjugate. For example, consider the decay $\bd \to K^0
\kstarbar$. As for the above processes, the amplitude can be written
\beq
A(\bd\to K^0\kstarbar) = \Ptwiduc\ e^{i \gamma} e^{i
{\tilde\delta}_{uc}} + \Ptwidtc\
e^{-i \beta} e^{i {\tilde\delta}_{tc}} ~.
\eeq
[The hadronic parameters are written with tildes to distinguish them
from their counterparts in Eq.~(\ref{BKKamp})]. For this decay, the
amplitude ${\bar A}$ for $\bdbar \to K^0 \kstarbar$ is not simply
related to that for $\bd \to K^0 \kstarbar$ since the hadronization is
different: in the latter decay, the spectator quark is part of the
$K^0$, while in the former it is contained in the $\kstarbar$. We
therefore write
\beq
A(\bdbar\to K^0\kstarbar) = \Ptwiduc' e^{-i \gamma} e^{i
{\tilde\delta}'_{uc}} +
\Ptwidtcp e^{i \beta} e^{i {\tilde\delta}'_{tc}} ~.
\eeq
By measuring $\bd(t)\to K^0\kstarbar$, one can obtain the observables
$X$, $Y$, $\ZI$, $\ZR$ defined previously. These now take the form
\bea
X & = & {1\over 2}
\left[ \Ptwiduc^2 + \Ptwidtc^2 - 2 \Ptwiduc \Ptwidtc \cos(\alpha -
{\tilde\Delta}) \right. \nn \\
& & \hskip1truecm
+ \left. {\tilde{\cal P}}_{uc}^{\prime^2} +  {\tilde{\cal
P}}_{tc}^{\prime^2} - 2 \Ptwiducp \Ptwidtcp \cos(\alpha +
{\tilde\Delta}') \right]
~, \nn \\
Y & = & {1\over 2}
\left[ \Ptwiduc^2 + \Ptwidtc^2 - 2 \Ptwiduc \Ptwidtc \cos(\alpha -
{\tilde\Delta}) \right. \nn \\
& & \hskip1truecm
- \left. {\tilde{\cal P}}_{uc}^{\prime^2} -  {\tilde{\cal
P}}_{tc}^{\prime^2} + 2 \Ptwiducp \Ptwidtcp \cos(\alpha +
{\tilde\Delta}') \right]
~, \nn \\
\ZI & = & \Ptwiduc \Ptwiducp \sin(2 \alpha - {\tilde\Delta} +
{\tilde\Delta}') \nn\\
& & \hskip2truemm - \Ptwiduc \Ptwidtcp \sin(\alpha - {\tilde\Delta})
- \Ptwidtc \Ptwiducp \sin(\alpha + {\tilde\Delta}') ~, \nn\\
\ZR & = &
\Ptwiduc \Ptwiducp \cos(2 \alpha - {\tilde\Delta} + {\tilde\Delta}') +
\Ptwidtc \Ptwidtcp \nn\\
& & \hskip2truemm - \Ptwiduc \Ptwidtcp \cos(\alpha - {\tilde\Delta})
- \Ptwidtc \Ptwiducp \cos(\alpha + {\tilde\Delta}') ~,
\label{BKK*obs1}
\eea
where ${{\tilde\Delta}}\equiv {{\tilde\delta}}_{uc} -
{{\tilde\delta}}_{tc}$ and
${{\tilde\Delta}}'\equiv {{\tilde\delta}}'_{uc} -
{{\tilde\delta}}'_{tc}$.

For the second process, it is natural to consider the conjugate final
state $\kbar K^*$. The amplitudes for $\bd$ and $\bdbar$ to decay to
this state are
\bea
A(\bd\to \kbar K^*) & = & \Ptwiducp e^{i \gamma} e^{i
{\tilde\delta}'_{uc}} + \Ptwidtcp
e^{-i \beta} e^{i {\tilde\delta}'_{tc}} ~, \nn\\
A(\bdbar\to \kbar K^*) & = &
\Ptwiduc\ e^{-i \gamma} e^{i {\tilde\delta}_{uc}} + \Ptwidtc\
e^{i \beta} e^{i {\tilde\delta}_{tc}} ~.
\eea
Measurements of $\bd(t)\to \kbar K^*$ then yield
\bea
X' & = & {1\over 2}
\left[ {\tilde{\cal P}}_{uc}^{\prime^2} +  {\tilde{\cal
P}}_{tc}^{\prime^2} - 2 \Ptwiducp \Ptwidtcp \cos(\alpha -
{\tilde\Delta}') \right. \nn \\
& & \hskip1truecm
+ \left. \Ptwiduc^2 + \Ptwidtc^2 - 2 \Ptwiduc \Ptwidtc \cos(\alpha +
{\tilde\Delta}) \right]
~, \nn \\
Y' & = & {1\over 2}
\left[ {\tilde{\cal P}}_{uc}^{\prime^2} +  {\tilde{\cal
P}}_{tc}^{\prime^2} - 2 \Ptwiducp \Ptwidtcp \cos(\alpha -
{\tilde\Delta}') \right. \nn \\
& & \hskip1truecm
- \left. \Ptwiduc^2 - \Ptwidtc^2 + 2 \Ptwiduc \Ptwidtc \cos(\alpha +
{\tilde\Delta}) \right]
~, \nn \\
\ZI' & = & \Ptwiduc \Ptwiducp \sin(2 \alpha + {\tilde\Delta} -
{\tilde\Delta}') \nn\\
& & \hskip2truemm - \Ptwiduc \Ptwidtcp \sin(\alpha + {\tilde\Delta})
- \Ptwidtc \Ptwiducp \sin(\alpha - {\tilde\Delta}') ~, \nn\\
\ZR' & = &
\Ptwiduc \Ptwiducp \cos(2 \alpha + {\tilde\Delta} - {\tilde\Delta}') +
\Ptwidtc \Ptwidtcp \nn\\
& & \hskip2truemm - \Ptwiduc \Ptwidtcp \cos(\alpha + {\tilde\Delta})
- \Ptwidtc \Ptwiducp \cos(\alpha - {\tilde\Delta}') ~.
\label{BKK*obs2}
\eea

As before, we have six independent observables as a function of seven
theoretical parameters, so we cannot obtain $\alpha$. However, one can
manipulate Eqs.~(\ref{BKK*obs1}) and (\ref{BKK*obs2}) to obtain
\beq
D' = C' \tan 2\alpha + {\Ptwidtcp\over\Ptwidtc} \, {B \over 2 \cos
2\alpha} -
{\Ptwidtc\over\Ptwidtcp} \, {B' \over 2 \cos 2\alpha} ~,
\label{alphasolve2}
\eeq
where
\bea
& B \equiv {1\over 2} (-X -Y + X' - Y') ~,~~
C' \equiv {1\over 2} (-\ZI + \ZI') ~, & \nn\\
& B' \equiv {1\over 2} (X -Y - X' - Y') ~,~~
D' \equiv {1\over 2} (\ZR - \ZR') ~.&
\eea
As in Eq.~(\ref{Pratios}) above, the ratio $\Ptwidtc/\Ptwidtcp$ can be
obtained from the ratio of branching ratios for $\bs\to K^0 \kstarbar$
and $\bs\to \kbar K^*$. Thus, Eq.~(\ref{alphasolve2}) can be used to
obtain $\alpha$, again up to discrete ambiguities.

{}From the above analysis, we therefore see that the CP phase $\alpha$
can be cleanly extracted from measurements of the decays of $\bd$ and
$\bs$ mesons to two different final states consisting of one neutral
kaon (i.e.\ $K^0$ or any of its excited states) and one neutral
anti-kaon (i.e.\ $\kbar$ or any excited state). However, note that the
$K^* {\bar K}^*$ final state actually consists of three helicity
states. Any of these can be considered a distinct final state for the
purposes of our analysis. Thus, by applying our method to two
different $K^* {\bar K}^*$ helicity states, $\alpha$ can be obtained
from $B^0_{d,s} \to K^* {\bar K}^*$ decays alone.

Of course, since they are pure $b\to d$ penguin decays, the branching
ratios for $\bd(t) \to K^{(*)} {\bar K}^{(*)}$ are expected to be
quite small, of order $10^{-6}$. Even so, since hadron colliders
produce an enormous number of $B$ mesons, such decays should be
measurable. Furthermore, in all cases, the kaon or anti-kaon can be
detected using its decays to charged $\pi$'s or $K$'s only; this
method does not require the detection of $\pi^0$'s. Therefore hadron
colliders will be able to use this technique to measure $\alpha$ --
all that is required is good $\pi$/$K$ separation. And if $\pi^0$'s
can be detected, this simply increases the detection efficiency for
the various final states.

Now, the key ingredient in the above method is the use of $\bs$ decays
to obtain information about the hadronic parameters of $\bd$ decays.
In Eq.~(\ref{Pratios}), we have assumed the equality of a double ratio
of matrix elements:
\beq
{r_t \over r_t^*} \equiv { \bra{K^0 \kbar} H_d \ket{\bd} / \bra{K^0
  \kbar} H_s \ket{\bs} \over \bra{K^* \kstarbar} H_d \ket{\bd} /
  \bra{K^* \kstarbar} H_s \ket{\bs} } = 1 ~,
\eeq
where we have defined $H_d \equiv (P_t - P_c)$ and $H_s \equiv
(P_t^{(s)} - P_c^{(s)})$. What is the error in making this assumption?
Consider first the ratio in the numerator, $r_t$. The two decays in
$r_t$ are related by U-spin, and so $r_t$ is equal to unity in the
chiral symmetry limit. Similar observations hold for the ratio in the
denominator, $r_t^*$.  We can therefore write
\bea
r_t & = & {\bra{K^0 \kbar} H_d \ket{\bd} \over \bra{K^0 \kbar} H_s
  \ket{\bs} } = 1 + C_{\sss SU(3)} ~, \nn\cr
r_t^* & = & { \bra{K^* \kstarbar} H_d \ket{\bd} \over \bra{K^*
\kstarbar} H_s \ket{\bs} } = 1 + C_{\sss SU(3)}^* ~,
\label{SU3break}
\eea
where $C_{\sss SU(3)}$ and $C_{\sss SU(3)}^*$ parametrize the size of
$SU(3)$ breaking in these ratios. Thus, we have
\beq
{r_t \over r_t^*} = 1 + (C_{\sss SU(3)} - C_{\sss SU(3)}^*) ~.
\eeq
Since there is no symmetry limit in which $(C_{\sss SU(3)} - C_{\sss
SU(3)}^*) \to 0$, apriori one would expect this quantity to be of
canonical $SU(3)$-breaking size, i.e.\ $O(25\%)$. However, as we argue
below, there are a number of reasons to expect significant
cancellations between $C_{\sss SU(3)}$ and $C_{\sss SU(3)}^*$.

We begin by examining the origin of $SU(3)$ breaking in $r_t$ alone.
First, consider the quark-level process underlying the $B^0_{d,s} \to
K^0 \kbar$ decays, ${\bar b} \to {\bar d} s {\bar s}$ or ${\bar b} \to
{\bar s} d {\bar d}$. The dominant configuration is the one in which
all three final-state quarks are energetic. Thus, in the limit $m_b
\to \infty$ we can neglect the masses of the light quarks, which
implies that, at the quark-level, $SU(3)$ breaking is negligible in
the decays $\bs \to K^0 \kbar$ and $\bd \to K^0 \kbar$. The
configuration in which one of the final-state quarks is soft is
suppressed by at least $1/\EK$ from the kaon wavefunction, where
$\EK=M_B/2$ is the energy of the final-state $K^0$ or $\kbar$.  In
addition, the annihilation contributions are suppressed by
$1/M_B$. Thus, the subdominant configurations are suppressed by a
factor of $1/M_B$ compared to the dominant one. Hence up to
corrections of $O([M_{B_d} - M_{B_s}]/M_B) \sim 2\%$, the hamiltonians
$H_d$ and $H_s$ are the same.

We can therefore write
\bea
r_t & = & \bra{K^0\kbar}H_d\ket{\bd}/\bra{K^0\kbar}H_s\ket{\bs} \nonumber\\
    & = & \bra{K^0\kbar}H_d\ket{\bd}/\bra{K^0\kbar}U^\dagger H_d U\ket{\bs} ~,
\label{ratio1}
\eea
where $U$ is the U-spin operator. Obviously, one would obtain $r_t=1$
if $SU(3)$ were a good symmetry, since then we would have $U\ket{\bs}
= \ket{\bd}$ and $U \ket{K^0\kbar} = \ket{K^0\kbar}$. However, $SU(3)$
is not a good symmetry, and this can affect $r_t$ in 2 distinct ways:
(i) ``final-state'' corrections, $U \ket{K^0\kbar} \ne
\ket{K^0\kbar}$, and (ii) ``initial-state'' corrections, $U\ket{\bs}
\ne \ket{\bd}$. In what follows we will examine in turn the size of
the $SU(3)$-breaking effects in each of these areas.

However, before doing so, we note that the sources of $SU(3)$
corrections in $r_t^*$ are very similar to those in $r_t$: $U \ket{K^*
\kstarbar} \ne \ket{K^* \kstarbar}$ and $U\ket{\bs} \ne \ket{\bd}$. It
is therefore not unreasonable to expect sizeable cancellations between
$C_{\sss SU(3)}$ and $C_{\sss SU(3)}^*$, leading to $r_t/r_t^* \simeq
1$.

We first consider $SU(3)$-breaking effects in the relation $U
\ket{K^0\kbar} = \ket{K^0\kbar}$. The wavefunction of an energetic $K^0$ or
$\kbar$ can be expanded in terms of Fock states as
\bea 
\psi_{\kbar} & =& \psi(s \bar{d}) +\psi(s \bar{d} g) + ... \nonumber\\ 
\psi_{K^0} & =& \psi(\bar{s} {d}) +\psi(\bar{s} {d} g) + ... 
\label{fock}
\eea
In general, the partons inside the energetic kaon are collinear and
have small transverse momentum. More precisely, the distribution in
the transverse momentum, $k_{\perp},$ is peaked at small values of
$k_{\perp} \sim \Lambda_{QCD}$ \cite{brodsky}. The contributions from
higher Fock states, in which the non-valence partons are hard and
carry a finite fraction of the kaon momentum, are suppressed by
$1/\EK$ because of the additional hard parton propagator in the final
state \cite{brodsky}. Hence we assume that the kaon wavefunction is
dominated by the valence-quark configuration. In this case, the
valence quarks each carry a certain fraction of the total kaon
momentum:
\bea
p_s & \approx & x p_K ~, \nonumber\\
p_d & \approx & (1-x) p_K ~,
\eea
with $0 \le x \le 1$.

However, note that, for the calculation of the nonleptonic amplitude,
what is relevant is not the full wavefunction of the kaon,
$\psi(x,k_{\perp})$, but rather its light cone distribution (LCD),
$\phi(x, \EK)$, which is related to the wavefunction by $\phi_K(x,\mu)
\sim \int^{\mu}\psi(x, k_{\perp}) d^2k_{\perp}$, where $\mu \sim \EK
\sim m_b$. Under a U-spin transformation the $s$ and $d$ quarks are
interchanged, so that
\beq
U\psi_K = \psi(d(x p_K) \bar{s}(1 - x p_K)) ~.
\label{ufock}
\eeq
{\it Thus, the U-spin breaking correction from the final state is due
to the presence of a piece in the kaon LCD which is antisymmetric
under the exchange $x \to 1-x$.}

Now, from QCD we know that the LCD's are {\it symmetric} under this
exchange as $\EK \to \infty$ \cite{brodsky}. Therefore, in the $\EK
\to \infty$ limit we have $U\ket{K^0}=\bar{\kbar}$ and
$U\bar{\kbar}=\ket{K^0}$. To be explicit, the leading-twist kaon LCD
$\phi_K(x,\mu)$ can be expanded in terms of Gegenbauer polynomials
$C^{3/2}_n$ as follows \cite{Ballv}:
\beq
\phi_K(x,\mu) = f_K \, 6x(1-x) \left( 1 + \sum_{n=1}^\infty
a_{2n}^K(\mu) C_{2n}^{3/2}(2x-1) \right) ~,
\eeq
where the Gegenbauer moments $a_n^K$ are multiplicatively
renormalized, change slowly with $\mu$, and vanish as $\mu \to
\infty$. It is the presence of the antisymmetric piece at scale $\mu
\sim m_b$, proportional to odd powers of $(2x - 1)$, which will
generate $SU(3)$ corrections from the final-state kaons.

Note that, in general, U-spin is {\it not} a good symmetry for final
states in the $E \to \infty$ limit. For example, it does not hold for
$K \leftrightarrow \pi$ transformations because the $K$ and $\pi$
wavefunctions are still different. Thus, for $K\pi$ final states, one
expects to obtain U-spin breaking effects of order $f_K/f_\pi$ in the
$E \to \infty$ limit.  However, U-spin {\it is} a good symmetry in the
$E \to \infty$ limit for $K^0 \leftrightarrow \kbar$ because the $K^0$
and $\kbar$ wavefunctions are the same. Thus, we see that $K^0\kbar$
is a {\it special final state} as far as U-spin (i.e.\ $SU(3)$) is
concerned.

We therefore see that SU(3) breaking in the final state is related to
the size of the antisymmetric piece of the kaon LCD at the scale of
$m_b$. However, there is indirect experimental evidence that this
antisymmetric piece may be absent: the recent measurement of the pion
LCD at $\mu^2\sim 10~{\rm GeV}^2$ \cite{pion} shows that the pion LCD
is extremely close to its asymptotic form, $\phi_\pi(x) \sim
x(1-x)$. (Note: isospin symmetry requires only that the pion LCD be
symmetric, not asymptotic.) This suggests that, at the scale $\mu \sim
m_b$, the LCD's of the light mesons $K$ and $K^*$ may also be very
close to their asymptotic form, i.e.\ symmetric under the interchange
$x \to 1-x$.  If this turns out to be the case, there would be tiny
$SU(3)$ breaking from the final states in the ratios $r_t$ and
$r_t^*$, and even tinier $SU(3)$ breaking in the quantity $r_t/r_t^*$.
Note that we only require the antisymmetric parts of the $K$ and $K^*$
LCD's to be absent in order to have tiny $SU(3)$ breaking in
$r_t/r_t^*$. (In fact, if the $K$ and $K^*$ LCD's were measured to be
symmetric, it would indicate that the difference between the $s$-quark
and $d$-quark masses is irrelevant for the $K$ and the $K^*$
wavefunctions.)

The main point here is that this can be tested experimentally: as was
done for the pion LCD, one can measure the LCD's of the $K$ and $K^*$
mesons. If they turn out to be symmetric, then, as was argued above,
the $SU(3)$-breaking contribution to $r_t$ and $r_t^*$ will be
negligible. On the other hand, if an antisymmetric piece is found,
then one needs to estimate its effect on the $B^0_{d,s} \to K^{(*)}
{\bar K}^{(*)}$ amplitudes. This requires a model calculation, and we
will come back to this possibility below.

We now turn to the initial-state corrections, due to $U\ket{\bs} \ne
\ket{\bd}$. In order to treat these quantitatively, we need a
framework in which to perform computations. QCD factorization
\cite{BBNS} is arguably the most developed calculational tool, so this
is what we shall use. Within QCD factorization, we write the ratio
$r_t$ schematically as
\beq
r_t = {A_{fac}^d \left[ 1 + x^d \right] \over A_{fac}^s \left[ 1 + x^s
\right] } ~,
\eeq
where $A_{fac}^d$ and $A_{fac}^s$ are the factorizable contributions
to $\bd \to K^0 {\bar K}^0$ and $\bs \to K^0 {\bar K}^0$,
respectively, and $x^d$ and $x^s$ are the corresponding
nonfactorizable contributions. Our definition of $A_{fac}^d$ and
$A_{fac}^s$ includes the $\alpha_s$ corrections to naive
factorization.

We first consider the nonfactorizable contributions.  An example of
such effects is the corrections due to hard gluon exchange between the
spectator quark and the energetic quarks of the emitted meson. These
have been calculated for $B\to \pi K$ decays \cite{BBNS}, and it has
been found that the $SU(3)$-breaking effects are quite small: their
size is given by
\beq
\left( {f_{B_s} \over \lambda_{B_s}} - {f_{B_d} \over \lambda_{B_d}}
\right) X ~,
\label{nonfac}
\eeq
where $M_{B_q}/\lambda_{B_q} = \int \phi_{B_q}(z)/z$, with
$\phi_{B_q}(z)$ being the $B^0_q$ LCD, $q=d,s$. The quantity $X$
depends on the final state. It is straightforward to adapt the
calculation of Ref.~\cite{BBNS} to $K {\bar K}$ final states: we find
that $(f_{B_d}/\lambda_{B_d}) X \lsim 10$\% (for $\pi K$ final states,
this quantity is $\lsim 5\%$). Now, in a simple model one can write
$f_{B_q} = \mu_q^{3/2}/M_{B_q}^{1/2}$ and $\lambda_{B_q} \sim \mu_q$,
where $\mu_q$ is the reduced mass, which is different for the $\bs$
and the $\bd$ mesons. Thus, in the heavy-quark limit we have
$f_{B_s}/f_{B_d} =\mu_s^{3/2}/\mu_d^{3/2}$ and
$(f_{B_s}/\lambda_{B_s}) / (f_{B_d}/\lambda_{B_d})
=\mu_s^{1/2}/\mu_d^{1/2}$. Taking $f_{B_s}/f_{B_d} = 1.15$, we find
that the $SU(3)$-breaking correction of Eq.~(\ref{nonfac}) is less
than 1\%. Thus, within QCD factorization, the $SU(3)$ corrections due
to nonfactorizable contributions are negligible. We will henceforth
concentrate only on the factorizable contributions $A_{fac}^d$ and
$A_{fac}^s$.

The factorizable contributions can be written as
\bea
A_{fac}^d = f_K \, \fbdk \, \int T(x) \phi_{\bar K}(x) dx ~, \nn\\
A_{fac}^s = f_K \, \fbsk \, \int T(x) \phi_K(x) dx ~.
\eea
In the above, $\fbdk$ and $\fbsk$ are form factors, while the
integrals represent  the hadronization of quarks into a $\kbar$ or
$K^0$. From the above, we see that there are two possible sources of
$SU(3)$ breaking: (i) the difference in the $\kbar$ and $K^0$ hadronization,
which is related to the difference the $\kbar$ and $K^0$ LCD's, and
(ii) the difference in form factors.

We first consider the $SU(3)$ breaking due to the $K^0$ and $\kbar$
LCD's. Since $\phi_{\bar K}(x) = \phi_K (1-x)$, it is clear that
$SU(3)$ breaking will only occur to the extent that the kaon LCD
contains an antisymmetric piece at the scale $\mu \sim m_b$. As has
already been discussed, if the kaon LCD turns out to be symmetric,
there are no final-state $SU(3)$ corrections to the amplitudes. This
is a model-independent result. However, even if $\phi_K(x)$ is found
to contain an antisymmetric piece, within QCD factorization it tends
not to contribute very much to the overall amplitude. For example, a
50\% asymmetry in the LCD of the kaon would only result in a $\sim
4\%$ $SU(3)$-breaking correction coming from the hard scattering part,
$\int T(x) \phi(x) dx $, for the $K^0 \kbar$ final state at the scale
$\mu=m_b$ \cite{BBNS}. (Note that the inclusion of an antisymmetric
piece of the kaon LCD introduces a scale dependence in the amplitude,
albeit at the $\alpha_s^2$ level. Thus, since $SU(3)$ corrections
cannot depend on the scale $\mu$, a proper $SU(3)$-breaking
calculation should include the full $\alpha_s^2$ calculation to the
nonleptonic amplitude.)

Furthermore, the final state consists of both a $K^0$ and $\kbar$.
Thus, if there is an antisymmetric piece in the kaon LCD, one would
expect some cancellation in the amplitude for $\bd \to K^0 \kbar$
between the hard scattering, which involves the $\kbar$, and the form
factor, which involves the $K^0$. A similar argument holds for $\bs
\to K^0\kbar$. In fact, the calculation of the $B_d \to K^0$ and $B_s
\to \kbar$ form factors using the same antisymmetric piece in the kaon
wavefunction results in about a 6\% final-state correction to the form
factors \cite{Chen}. This is partially cancelled by the 4\% correction
coming from the hard scattering part, resulting in an $SU(3)$
correction of $\sim 2\%$ in $r_t$. Note: since the approach to
nonleptonic decays in Ref.~\cite{Chen} (perturbative QCD) is slightly
different than that of Ref.~\cite{BBNS} (QCD factorization), one has
to be careful about combining their results. However, it is reasonable
to expect that the net $SU(3)$ breaking will be only about a few
percent, as estimated above. A similar analysis holds for the $K^*$
final state and the ratio $r_t^*$. In addition, the $SU(3)$ breaking
in the $K$ and the $K^*$ LCD's is very similar: model calculations
\cite{Ballv} find the antisymmetric pieces to be equal at the level of
$\lsim 10\%$. This indicates that the small $SU(3)$ breaking from the
final-state mesons in $r_t$ and $r_t^*$ will come with the same sign
and will partially cancel in the ratio $r_t/r_t^*$.

We therefore see that the various model calculations lead to the
conclusion that the final-state $SU(3)$ breaking in $r_t/r_t^*$ is
tiny even in the presence of a sizeable antisymmetric piece in the $K$
and the $K^*$ LCD's. Thus, although we admittedly can give no formal
proof of this, it seems quite likely that the $SU(3)$ breaking in the
final state is indeed very small.

{}From the above analysis, it appears that the main contribution to
$SU(3)$ breaking in $r_t$ (and $r_t/r_t^*$) comes from the $B\to K$
form factors. However, QCD factorization says nothing about how to
calculate these quantities. Fortunately, we can use experimental
measurements to obtain information about the form factors. The main
observation is that $B\to K$ form factors are related to $D\to K$ form
factors\footnote{We thank C. Bauer, D. Pirjol and I. Stewart for
pointing this out to us.}. In particular, we note that in the chiral
limit and in the heavy-quark limit \cite{Grinstein},
\beq
{ \fbdk / \fbsk \over \fdk / \fdsk} = 1 ~.
\eeq
We therefore conclude that the deviation of this quantity from unity
is at most $O([M_D - M_{D_s}]/M_D) \sim 5\%$. In other words, the
measurement of the ratio of $D \to K$ form factors at $q^2 = 0$ (for
example, in semileptonic $D$ decays) will indirectly give us the ratio
of the $B\to K$ form factors at $q^2 = 0$, up to $O(5\%)$ corrections.
(There is a slight subtlety here: $q^2 = 0$ for $D\to K$ form factors
corresponds to a kaon energy $\EK = M_D/2$, whereas $q^2 = 0$ for
$B\to K$ form factors implies a larger value of the kaon energy: $\EK
= M_B/2$. Thus, if the measurement of $\fdk/\fdsk$ yields a deviation
from 1 of $X\%$, $\fbdk/\fbsk - 1$ is expected to be {\it less than
X\%}.)

In fact, this relation between the $B\to K$ and $D\to K$ form factors
may allow us to deduce that the initial-state $SU(3)$-breaking
corrections are absent: if it is found experimentally that $\fdk/\fdsk
\simeq 1$, then this implies that $\fbdk/\fbsk \simeq 1$, so that the
$SU(3)$-breaking correction of Eq.~(\ref{SU3break}) is $C_{\sss SU(3)}
\simeq 0$. If a similar result is found for the $D\to K^*$ form
factors, one will conclude that $C_{\sss SU(3)}^* \simeq 0$ as well.
It is therefore possible to establish experimentally that the $SU(3)$
corrections in $r_t$, $r_t^*$ and $r_t/r_t^*$ are small.

Suppose instead that the ratio $\fdk/\fdsk$ is found to deviate from
unity, and similarly for the $D\to K^*$ form factors. We can therefore
write
\bea
{ \fbdk / \fbsk \over \fdk / \fdsk} & = & 1 + a \, {\Delta M_D \over M_D}
~, \nn\\
{ \fbdkstar / \fbskstar \over \fdkstar / \fdskstar}  &=& 1 + a^* \, {\Delta
M_D \over M_D} ~, 
\eea
where $a$ and $a^*$ are numbers of $O(1)$. That is,
\beq
{\fbdk / \fbsk \over \fbdkstar / \fbskstar}  = {\fdk / \fdsk \over \fdkstar
/ \fdskstar} \, \left[ 1 + (a - a^*) {\Delta M_D \over M_D} \right] ~.
\label{ddratio}
\eeq
In other words, the measurement of the $D \to K$ and $D\to K^*$ form
factors determines the relevant ratio of $B\to K$ and $B\to K^*$ form
factors up to corrections of $O(5\%)$. Thus, this gives us a method of
experimentally measuring the $B$ form factors.

Furthermore, because the various $B$ decays are so similar, one might
expect that $a \simeq a^*$ in the above relation, so that the
correction to the ratio of ratios of $B$ form factors is in fact
smaller than 5\%. This is indeed the case: for the pseudoscalar-vector
final state, model calculations give $(a-a^*) (\Delta M_D/M_D) < 1\%$
\cite{stech}.

We note in passing that there are other experimental measurements
which probe the size of initial-state $SU(3)$ breaking. For example,
neglecting the OZI-suppressed penguin contribution, one expects
\beq
{\Gamma (\bs\to \Psi \ks) \over \Gamma (\bd\to\Psi \ks)} = \left|
{V_{cd} \over V_{cs}} \right|^2 \left( 1 + {\hbox{$SU(3)$ breaking}}
\right).
\eeq
Thus, if these two rates are measured to be equal, up to the ratio of
CKM factors, this will support the conjecture that initial-state
$SU(3)$-breaking effects in $B\to K$ transitions are in fact rather
small.

Finally, we have information about the $B\to K$ and $B\to K^*$ form
factors in the limit of $m_b \to \infty$ and $\EK \to \infty$. The
authors of Ref.~\cite{Charles} showed that in this limit only three
form factors, $\xi, \xi_{\|}$ and $\xi_{\perp}$, are necessary to
describe $B\to K$ and $B\to K^*$ semileptonic transitions. They went
on to calculate the form factors using QCD sum rules. This approach
reproduces the symmetry relations among form factors in the $m_b,\EK
\to \infty$ limit. The three form factors are given by
\bea
\xi &=& \frac{1}{f_B}\frac{1}{2E^2}[-f_P
\phi^{\prime}(1)I_2(\omega_0,\mu_0) +f_P \frac{m_P^2}{m_s+m_d}
\phi_p(1)I_1(\omega_0,\mu_0)] ~, \nonumber\\
\xi_{\|} &=& \frac{1}{f_B}\frac{1}{2E^2}[-f_V
\phi^{\prime}_{\|}(1)I_2(\omega_0,\mu_0)
+f_V^{\perp}{m_V}h^{t}_{\|}(1)I_1(\omega_0,\mu_0)] ~, \nonumber\\
\xi_{\perp} &=& \frac{1}{f_B}\frac{1}{2E^2}[-f_V^{\perp}
\phi^{\prime}_{\perp}(1)I_2(\omega_0,\mu_0)
+f_V{m_V}g^{v}_{\perp}(1)I_1(\omega_0,\mu_0)] ~,
\label{qcdsumrule}
\eea
where $f_B,f_P,f_V$ and $f_V^{\perp}$ are decay constants, and
$\phi,\phi_p, \phi_{\|}, \phi_{\perp},h^t_{\|}$ and $g^v_{\perp}$ are
the asymptotic twist-2 and twist-3 LCD's of the $K$ and the $K^*$. All
the initial-state effects are contained in $f_B$ and the integrals
$I_{1,2}(\omega_0, \mu_0)$, which are given by
\beq
I_{i}(\omega_0, \mu_0) = exp[\frac{2\Lambda}{\mu_0}]\int_0^{\omega_0}
{d \omega \omega^i exp[-\frac{2\omega}{\mu_0}]} ~,
\eeq
where $\Lambda_q=M_{B_q}-m_b$. Note that $\Lambda_d$ and $\Lambda_s$
are in general different and will generate initial-state $SU(3)$
breaking. The parameters $\omega_0$ and $\mu_0$ of the model are
defined in Refs.~\cite{Ballv,Charles} and can be taken to be the same
for $\bd$ and $\bs$ decays \cite{Ballv}, where a tiny flavor
dependence for $\omega_0$ has been neglected. It is then
straightforward to see that Eq.~(\ref{qcdsumrule}) implies that
$(\fbdk / \fbsk) / (\fbdkstar / \fbskstar) = 1$, i.e.\ that all
initial-state effects cancel. There are, in principle,
$SU(3)$-breaking corrections to the form factors due to corrections of
$O(\alpha_s)$. However, these corrections are themselves very small at
$\EK =M_B/2$ \cite{feldmann}, and consequently the $SU(3)$ breaking
from them is totally negligible.  Thus, within the QCD sum rule
approach, if the LCD's of the $K$ and the $K^*$ are symmetric at the
$m_b$ scale, one has $(\fbdk / \fbsk) / (\fbdkstar / \fbskstar) =
1$. Then, using QCD factorization one deduces that $r_t/r_t^* =1$, up
to corrections of $O([M_{B_d} - M_{B_s}]/M_B)$.

To summarize the above discussion: our method assumes the equality of
$r_t/r_t^*$, a double ratio of $B_{d,s} \to K^{(*)} {\bar K}^{(*)}$
matrix elements. This ratio can deviate from unity due to flavor
$SU(3)$-breaking effects. Some of these effects (e.g. corrections to
the hamiltonian, annihilation contributions, etc.) can be shown to be
suppressed by $1/M_B$, and so are expected to be at most $O(\Delta M_B
/ M_B) \simeq 2\%$. The potentially large corrections are due to
final-state effects ($U \ket{K^0\kbar} \ne \ket{K^0\kbar}$) and
initial-state effects ($U\ket{\bs} \ne \ket{\bd}$). Although we cannot
formally prove that these effects are small, all model calculations
suggest this to be the case. Furthermore, there are a variety of
experimental measurements which can test this conclusion. Taking all
the model calculations into account, our best guess is that
$SU(3)$-breaking effects cause $r_t/r_t^*$ to deviate from unity by at
most $5\%$, and it would not be at all suprising if this deviation
turns out to be even smaller, say $\lsim 1\%$.

There is one other source of theoretical uncertainty: in
Eq.~(\ref{Bsdecay}), we have neglected the $(P_u^{(s)} - P_c^{(s)})
V_{ub}^* V_{us}$ term compared to $(P_t^{(s)} - P_c^{(s)}) V_{tb}^*
V_{ts}$. The justification is principally the size of the CKM matrix
elements: we have $|(V_{ub}^* V_{us})/(V_{tb}^* V_{ts})| \simeq
|V_{us}| |V_{ub}/V_{cb}| \simeq 0.02$, where we have taken
$|V_{ub}/V_{cb}| = 0.09$ \cite{PDG}. However, there is also a
suppression from the penguin matrix elements: for $\bd$ decays,
$|P_u|$ and $|P_c|$ are expected to be at most 50\% of $|P_t|$, and
are probably smaller. (For example, in Ref.~\cite{DKL} it is found
that, for $\bd \to K^* \kstarbar$, $0.14 \le |P_c-P_u| / |P_t-P_u| \le
.54$.) As argued above, this will not change significantly for $\bs$
decays. We therefore conclude that the error made in neglecting the
$(P_u^{(s)} - P_c^{(s)}) V_{ub}^* V_{us}$ term in Eq.~(\ref{Bsdecay})
is less than 1\%.

We now turn to an examination of the discrete ambiguities inherent in
this method. Consider the pair of decays $\bd \to K^0 \kbar$ and $\bd
\to K^* \kstarbar$. Let us assume that the true values of the
theoretical parameters are
\bea
& \Ptc = 1.1 ~,~~ \Puc = 0.4 ~,~~
\Ptc' = 1.3 ~,~~ \Puc' = 0.2 ~, & \nn\\
& \Delta = 40^\circ ~,~~ \Delta' = 70^\circ ~,~~ \alpha = 110^\circ ~. &
\label{input1}
\eea
Given these inputs, we can calculate the values of the experimental
quantities in Eqs.~(\ref{BKKobservables}) and (\ref{ZR}), as well as
their primed counterparts. Then, assuming that $\Ptc^2 / {\Ptc'}^2$
has been obtained from the decays $\bs \to K^0 \kbar$ and $\bs \to K^*
\kstarbar$ as in Eq.~(\ref{Pratios}), we can use
Eq.~(\ref{alphasolve}) to obtain $\alpha$.

\begin{table}
\hfil
\vbox{\offinterlineskip
\halign{&\vrule#&
 \strut\quad#\hfil\quad\cr
\noalign{\hrule}
height2pt&\omit&&\omit&&\omit&&\omit&&\omit&&\omit&&\omit&\cr
& $\alpha$ && $\Ptc$ && $\Puc$ && $\Delta$ && $\Ptc'$ && $\Puc'$ &&
$\Delta'$ & \cr
height2pt&\omit&&\omit&&\omit&&\omit&&\omit&&\omit&&\omit&\cr
\noalign{\hrule}
height2pt&\omit&&\omit&&\omit&&\omit&&\omit&&\omit&&\omit&\cr
& $110^\circ$ && 1.1 && 0.4 && $40^\circ$ && 1.3 && 0.2 && $70^\circ$ &
\cr
& $160^\circ$ && 3.0 && 3.5 && $175.8^\circ$ && 3.6 && 3.9 &&
$177.2^\circ$ & \cr
& $34.9^\circ$ && 1.7 && 0.7 && $25^\circ$ && 2.0 && 2.3 && $5^\circ$ &
\cr
& $55.1^\circ$ && 1.2 && 1.5 && $10.9^\circ$ && 1.4 && 0.2 &&
$62.2^\circ$ & \cr
& $101.8^\circ$ && 0.2 && 1.2 && $76.6^\circ$ && 0.3 && 1.4 &&
$139.7^\circ$ & \cr
& $168.2^\circ$ && 1.1 && 1.8 && $139.2^\circ$ && 1.3 && 0.9 &&
$107.6^\circ$ & \cr
& $131.9^\circ$ && 1.0 && 0.5 && $47.8^\circ$ && 1.1 && 1.8 &&
$171.2^\circ$ & \cr
& $138.1^\circ$ && 1.1 && 1.8 && $168.1^\circ$ && 1.3 && 0.3 &&
$76^\circ$ & \cr
height2pt&\omit&&\omit&&\omit&&\omit&&\omit&&\omit&&\omit&\cr
\noalign{\hrule}}}
\caption{
Solutions for $\alpha$ [from Eq.~(\protect\ref{alphasolve})] and
hadronic quantities, from measurements of $\bd \to K^0 \kbar$ and $\bd
\to K^* \kstarbar$, assuming the input values given in
Eq.~(\protect\ref{input1}).
}
\label{output1}
\end{table}

The results are shown in Table \ref{output1}. There are a total of 16
solutions for $\alpha$: in addition to the 8 solutions shown in the
Table, solutions with $\alpha \to \alpha + \pi$ are also allowed if
one simultaneously takes $\Delta \to \Delta + \pi$ and $\Delta' \to
\Delta' + \pi$ as well. This large number of discretely ambiguous
solutions for $\alpha$ is potentially a serious drawback of this
method. However, there are two ways of reducing the discrete
ambiguity.

First, one can also consider a different pair of $K^{(*)} {\bar
  K}^{(*)}$ final states. In this case one expects that the hadronic
quantities will take very different values. Because of this, although
one still expects a large number of possible solutions for $\alpha$,
these solutions will, in general, be different from those found in
Table \ref{output1}.

This is indeed what happens. For example, consider now the pair of
decays $\bd \to K^0 \kstarbar$ and $\bd \to \kbar K^*$, and assume
that the hadronic input values are
\bea
& \Ptwidtc = 1.2 ~,~~ \Ptwiduc = 0.2 ~,~~
\Ptwidtcp = 1.0 ~,~~ \Ptwiducp = 0.3 ~, & \nn\\
& {\tilde\Delta} = 80^\circ ~,~~ {\tilde\Delta}' = 120^\circ ~. &
\label{input2}
\eea
(Of course, $\alpha$ is assumed to take the same value as in
Eq.~(\ref{input1}), $110^\circ$.) As before, we use these input
quantities calculate the values of the observables, and we then solve
Eq.~(\ref{alphasolve2}) to obtain $\alpha$.

The results are shown in Table \ref{output2}. As before, we show only
8 solutions for $\alpha$; there are another 8 solutions with $\alpha
\to \alpha + \pi$. However, a comparison of Tables \ref{output1} and
\ref{output2} reveals that only two of the eight solutions are common
to both sets of processes: $110^\circ$ (the true solution) and
$160^\circ$. Thus, by applying the method to several sets of final
states, one can reduce the ambiguity in $\alpha$ to a fourfold one.

\begin{table}
\hfil
\vbox{\offinterlineskip
\halign{&\vrule#&
 \strut\quad#\hfil\quad\cr
\noalign{\hrule}
height2pt&\omit&&\omit&&\omit&&\omit&&\omit&&\omit&&\omit&\cr
& $\alpha$ && $\Ptwidtc$ && $\Ptwiduc$ && ${\tilde\Delta}$ &&
$\Ptwidtcp$ && $\Ptwiducp$ && ${\tilde\Delta}'$ & \cr
height2pt&\omit&&\omit&&\omit&&\omit&&\omit&&\omit&&\omit&\cr
\noalign{\hrule}
height2pt&\omit&&\omit&&\omit&&\omit&&\omit&&\omit&&\omit&\cr
& $110^\circ$ && 1.2 && 0.2 && $80^\circ$ && 1.0 && 0.3 && $120^\circ$ &
\cr
& $160^\circ$ && 3.3 && 3.6 && $176.8^\circ$ && 2.8 && 2.8 &&
$174.6^\circ$ & \cr
& $42^\circ$ && 0.5 && 1.5 && $25.9^\circ$ && 0.4 && 0.9 && $95.3^\circ$
& \cr
& $48^\circ$ && 0.5 && 0.9 && $135.8^\circ$ && 0.4 && 1.1 &&
$53.4^\circ$ & \cr
& $15.8^\circ$ && 4.1 && 4.4 && $2.6^\circ$ && 3.4 && 3.5 && $4.3^\circ$
& \cr
& $74.2^\circ$ && 1.2 && 0.3 && $126.3^\circ$ && 1.0 && 0.3 &&
$79.9^\circ$ & \cr
& $132.5^\circ$ && 0.5 && 0.9 && $42.4^\circ$ && 0.4 && 1.1 &&
$133.9^\circ$ & \cr
& $137.5^\circ$ && 0.5 && 1.5 && $156.4^\circ$ && 0.5 && 0.8 &&
$78.3^\circ$ & \cr
height2pt&\omit&&\omit&&\omit&&\omit&&\omit&&\omit&&\omit&\cr
\noalign{\hrule}}}
\caption{
Solutions for $\alpha$ [from Eq.~(\protect\ref{alphasolve2})] and
hadronic quantities, from measurements of $\bd \to K^0 \kstarbar$ and
$\bd \to \kbar K^*$, assuming the input values given in
Eq.~(\protect\ref{input2}).
}
\label{output2}
\end{table}

The second way to reduce the discrete ambiguity is to use the fact
that, as discussed above, we expect each of $\Puc/\Ptc$,
$\Puc'/\Ptc'$, $\Ptwiduc/\Ptwidtc$ and $\Ptwiducp/\Ptwidtcp$ to be
less than about 0.5 in the SM. This constraint eliminates most of the
solutions in Tables \ref{output1} and \ref{output2}. In fact, by
combining both methods, one can measure $\alpha$ with only a twofold
ambiguity: $\{ \alpha, ~\alpha + \pi \}$. Unless one has knowledge
about the strong phases, this discrete ambiguity cannot be further
reduced.

Finally, as we have argued above, this method for measuring $\alpha$
includes a theoretical uncertainty of at most 5\%. How does this error
quantitatively affect the extraction of $\alpha$? One can compute this
by allowing $\Ptc/\Ptc'$ and $\Ptwidtc/\Ptwidtcp$ to vary by $\pm 5\%$
in Eqs.~(\ref{alphasolve}) and (\ref{alphasolve2}). For the particular
cases considered above [Eqs.~(\ref{input1}) and (\ref{input2})], we
find that the theoretical uncertainty leads to an error on $\alpha$ of
$\pm 12^\circ$. On the other hand, if the theoretical error can be
shown to be smaller, say 1\%, then the error on $\alpha$ is reduced
considerably to $\pm 2^\circ$. Furthermore, for other choices of input
parameters, the error on $\alpha$ can be even smaller. This occurs
when the hadronic quantities describing the two final states are very
different. Thus, the method is most accurate when two very dissimilar
final $K^{(*)} {\bar K}^{(*)}$ states are used.

In summary, we have presented a new method for measuring $\alpha$.  It
involves the measurements of $\bd$ and $\bs$ decays to $K^{(*)} {\bar
K}^{(*)}$ final states. The method is very clean: based on a variety
of model calculations, we estimate that the theoretical uncertainty is
at most 5\%, and it would not be surprising if it turned out to be
even smaller. Furthermore, there are several experimental measurements
which can be used to probe the size of the theoretical error. Although
there are multiple discrete ambiguities in the extraction of $\alpha$,
by applying the method to several different final states, it is
possible to obtain $\alpha$ with a fourfold ambiguity. If an
additional (justified) assumption is made, the ambiguity can be
reduced to twofold: $\{ \alpha, ~\alpha + \pi \}$. Since this method
does not require $\pi^0$ detection, it is appropriate for use at
hadron colliders.

\bigskip
\noindent
{\bf Acknowledgements}:
We are grateful to A. Petrov, S. Brodsky, C. Bauer, D. Pirjol and
I. Stewart for discussions about the size of $SU(3)$ breaking in the
$B^0_{d,s} \to K^0 {\bar K}^0$ amplitudes. We thank D. Pirjol for
comments on the manuscript. This work was financially supported by
NSERC of Canada.


\end{document}